\documentclass[aps,twocolumn,prl]{revtex4}
\usepackage{graphicx}
\usepackage[]{epsfig}
\begin{document}


\noindent {\bf Comment on ``Theoretical Analysis of the
Transmission Phase Shift of a Quantum Dot in the Presence of Kondo
Correlations"}

\vspace{.5cm}
 In a recent Letter \cite{lavagna},
Jerez, Vitushinsky and Lavagna (JVL) propose a new theoretical
interpretation of the experimental measurements \cite{heiblum} of
the transmission phase shift, $\delta_{ABI}$, through a quantum
dot (QD) in the Kondo regime, as deduced from placing the QD in an
open double-slit Aharonov-Bohm interferometer (ABI). Describing
the QD, which is coupled to two reservoirs via one-dimensional
leads, by the single level Anderson model (SLAM) [their Eq.~(1)],
JVL argue that at zero magnetic field ($H=0$) the conductance
through the QD is given by $G \propto \sin^2(\delta_G)$, with
$\delta_G=\delta_{ABI}/2$.
 In this Comment we question the
validity of this result for the SLAM, since it fails in several
exactly known limits. Whether the SLAM succeeds to describe the
experiment~\cite{heiblum} is irrelevant to the theoretical problem
posed here~\cite{AA1}.

  Without interactions
($U=0$), JVL's SLAM can be solved exactly: assuming the scattering
solution $A_{\ell}e^{ikx}+B_{\ell}e^{-ikx}$ to the left of the QD,
and $A_{r}e^{-ikx}+B_{r}e^{ikx}$ to its right, the solution is
\begin{eqnarray}
\left[\begin{array}{c}B_{\ell}\\B_{r}\end{array}\right ]=S_{k}
\left[\begin{array}{c}A_{\ell}\\A_{r}\end{array}\right ],
\end{eqnarray}
with $E_{k}=-2t\cos k$ (we use all the notations of Ref.~
\cite{lavagna}) and with the scattering matrix given by
\begin{eqnarray}
&&S_{k}=\left [\begin{array}{cc}-1+2i\sin
k~ {\cal G}_{k}V_{L}^{2}/t&2i\sin k ~{\cal G}_{k}V_{L}V_{R}/t\\
2i\sin k~ {\cal G}_{k}V_{L}V_{R}/t&-1+2i\sin k ~{\cal
G}_{k}V_{R}^{2}/t\end{array}\right ]\nonumber\\
&&\equiv-e^{i\delta_{k}}\left [\begin{array}{cc}
\cos\delta_{k}-iv_{-}\sin\delta_{k}
&iv_{+}\sin\delta_{k}\\
iv_{+}\sin\delta_{k}& \cos\delta_{k}+iv_{-}\sin\delta_{k}
\end{array}\right ],\label{scat1}
\end{eqnarray}
where ${\cal G}_{k}=\bigl [E_{k}-\epsilon_{0}+e^{ik}\Gamma/2\bigr
]^{-1}$, with $\Gamma = 2 (V_{L}^{2}+V_{R}^{2})/t$, ${\rm
cot}\delta_{k}=-[E_{k}-\epsilon_{0}+(\Gamma/2) \cos k]
/[(\Gamma/2) \sin k ]$, while $v_{+}=\sin 2\theta $ and
$v_{-}=\cos 2\theta$, with $\tan\theta =V_{L}/V_{R} $. When $H=0$,
 $S_k$ is the same for both spin
indices, and we thus omit the spin index $\sigma$.

JVL indeed write down expressions which are equivalent to our
Eq.~(\ref{scat1}) (with $V_L=V_R$, i.e. $v_+=1,~v_-=0$), but they
then replace this equation (at the Fermi level, $k=k_F$) by their
Eq.~(4),
\begin{eqnarray}
S^{JVL}_{k_F\sigma}=e^{i\delta}\left [\begin{array}{cc}
\cos\delta_{\sigma}
&i\sin\delta_{\sigma}\\
i\sin\delta_{\sigma}& \cos\delta_{\sigma}
\end{array}\right ],
\label{JVL}
\end{eqnarray}
with the modified overall phase
$\delta=\delta_\uparrow+\delta_\downarrow$ (and a different sign).
This amounts to multiplying Eq.~(\ref{scat1}) by an additional
factor, $-C_\sigma=-e^{i\delta_{-\sigma}}$, which JVL attribute to
generalizations of Levinson's theorem. Although the conductance is
still given by $G \propto \sum_\sigma \sin^2\delta_\sigma$, the
ABI phase $\delta_{ABI}$ is then claimed to be equal to $\delta$.
 For $H=0$, one has
$\delta_\uparrow=\delta_\downarrow=\delta/2$, and therefore JVL
conclude that $\delta_G=\delta/2=\delta_{ABI}/2$.

However, for $U=0$  Eq.~(\ref{JVL}) {\it contradicts the exact
solution} (\ref{scat1}), which does {\it not} contain the factor
$-C_\sigma$.
 More generally, at zero-temperature but $U \ne 0$, one has $G \propto
\sum_\sigma \text{Im} {\cal G}_{d \sigma} (0) \propto \sum_\sigma
\sin^2 \delta_\sigma$, where ${\cal G}_{d \sigma} (0) \equiv e^{i
\delta_{\sigma}}\left|{\cal G}_{d \sigma} (0)\right|$ is the exact
local Green's function of the SLAM at the Fermi
energy~\cite{NG,Gerland}.
 Moreover,  Eq.~(2) of \cite{Gerland} shows generally that the complex transmission
amplitude $T_{d\sigma}$ through a SLAM QD is proportional to
${\cal G}_{d \sigma} (0)$, implying that $\delta_{ABI}= \arg
T_{d\sigma} \equiv \delta_\sigma$ is {\it the same} as $\delta_G$,
again contradicting JVL's $\delta_G=\delta_{ABI}/2$~\cite{AA}.

We conclude that JVL's Eq.~(4) does {\it not} follow from the
SLAM. Either the SLAM is not compatible with the Levinson theorem,
or the application of this theorem to the SLAM was done
incorrectly. In either case, if JVL believe that their Eq.~(4) is
correct then they should supply its explicit derivation  from a
well defined model.

We acknowledge clarifying correspondence with M. Lavagna. This
work  is supported by an ISF center of excellence (TAU),
by ISF, 
Minerva and BSF (WIS), by DIP (WIS and LMU), and by DFG, SFB631 (LMU).

 \vspace{.2cm}

A. Aharony and  O. Entin-Wohlman

{\it Department of Physics, Ben Gurion University, Beer Sheva 84105,
Israel, and School of Physics and Astronomy, Tel Aviv University,
Tel Aviv 69978, Israel}

Y. Oreg

{\it Department of Condensed Matter Physics, The Weizmann Institute
of Science, Rehovot 76100, Israel}

J. von Delft

{\it Physics Department, ASC and CeNS,
Ludwig-Maximilians-Universit\"at M\"unchen, D-80333 M\"unchen}

\vspace{.2cm}

\end{document}